\documentclass[conference]{IEEEtran} 
\usepackage{fancyhdr}
\usepackage{epsfig}
\usepackage{threeparttable}
\usepackage{epsf,epsfig}
\usepackage{amsthm}
\usepackage{amsmath}
\usepackage{amssymb}
\usepackage{amsfonts}
\usepackage[noadjust]{cite}
\usepackage{dsfont}
\usepackage{subfigure}
\usepackage{color}
\usepackage{enumerate}

\newtheorem{corollary}{Corollary}

\newtheorem{proposition}{Proposition}

\def\E{\mathsf{E}}

\def\SIR{\mathsf{SIR}}

\def\l{\left}
\def\r{\right}
\def\({\left(}
\def\){\right)}

\def\[{\left[}
\def\]{\right]}

\def\papertitle{Asymptotic Behavior of Ultra-Dense Cellular Networks and Its Economic Impact}

\IEEEoverridecommandlockouts

\begin{document}

\title{ \fontsize{23}{24}\selectfont \papertitle}


\author{Jihong Park, Seong-Lyun Kim, and Jens Zander
\thanks{J. Park and S.-L. Kim are with Dept. of Electrical \& Electronic Engineering, Yonsei University, Seoul, Korea (email: jhpark.james@yonsei.ac.kr, slkim@yonsei.ac.kr). J. Zander is with Wireless@KTH, KTH -- The Royal Institute of Technology, Stockholm, Sweden (email: jensz@kth.se). }}

\maketitle

\begin{abstract} 
This paper investigates the relationship between base station (BS) density and average spectral efficiency (SE) in the downlink of a cellular network. This relationship has been well known for sparse deployment, i.e. when the number of BSs is small compared to the number of users. In this case the SE is independent of BS density. As BS density grows, on the other hand, it has previously been shown that increasing the BS density increases the SE, but no tractable form for the SE-BS density relationship has yet been derived. In this paper we derive such a closed-form result that reveals the SE is asymptotically a logarithmic function of BS density as the density grows. Further, we study the impact of this result on the network operator's profit when user demand varies, and derive the profit maximizing BS density and the optimal amount of spectrum to be utilized in closed forms. In addition, we provide deployment planning guidelines that will aid the operator in his decision if he should invest in densifying his network or in acquiring more spectrum.

\end{abstract}

\section{Introduction}

Deploying more cellular base stations (BS) has been the main remedy to cope with relentless traffic growth. To reach the highest capacities, interest has lately been turning toward ultra-dense BS deployment \cite{Holistic13}, \cite{Zander13}, where the density of BS may even exceed the typical number of users in a given area. The impact of such extreme BS densification, however, has still not been explicitly analyzed.

To be more specific, in an engineering perspective, the preceding work \cite{Andrews:2011bg} provides an analytic average spectral efficiency (SE) calculation that reveals the SE is independent of BS density. The result is accurate only when BS density is low as it relies on an assumption that every BS has at least a single serving user. For densely deployed BS environment, the authors \cite{Yu2011, SLeeKHuang12,Alexiou13} consider turned-off BSs when having no serving users as in the Third Generation Partnership Project (3GPP) Release 12 specifications \cite{NCT2013}, and predict the SE is a logarithmic function of BS density via its compatible performance metrics, service capacity, outage probability, and common (worst user's) rate respectively. Nevertheless, they cannot represent the explicit relationship between the SE and BS density due to the intractable forms of results.

Motivated by these discussions on the BS densification, we derive an analytic SE expression valid for general BS density by utilizing a stochastic geometric approach \cite{StoyanBook:StochasticGeometry:1995}, and further provide its closed-form representation under ultra-dense BS environment. The result verifies \emph{SE logarithmically increases with BS density} in ultra-dense networks where the BS densification makes interference restricted by turning off empty BSs.

In an economic perspective, the authors \cite{JHuang11} consider a network operator's profit maximization based on user demand prediction in a single cell scenario neglecting the BS densification effect. For a multi-cell network, the work \cite{SLeeKHuang12} deals with cost minimization when BS density is low. Regarding the relationship between user demand and network supply, the authors \cite{Ericsson12} provide an analytic approach although it resorts to an iterated simulation.

This study, leading from the preceding works, focuses on the question: How much amount of BS density and spectrum a network operator should invest in when user demand changes? To capture the user demand variation, we consider two user demand characteristics: (i) user density and (ii) each user's sensitivity to his downloading rate. We thereby answer the question via solving a profit maximization problem while considering the user demand meeting network's rate supply. 

This paper examines the ramifications of BS density increase in downlink cellular networks from both engineering and economic points of view in terms of SE and profit respectively. The main contributions are listed below.
\begin{enumerate}[  1.]
\item This paper derives the closed-form SE in an ultra-dense downlink cellular network that is a logarithmic function of BS density.
\item The paper provides network planning guidelines in terms of the profit optimal BS density and spectrum amount in closed forms. 
\end{enumerate}

\section{Average Spectral Efficiency in Downlink Cellular Networks} \label{Sect:Cap}

\subsection{Network Model} \label{Sect:Cap_NetModel}
We consider a downlink cellular network. Let $\Phi_b$ denote BS coordinates in a two-dimensional Euclidean plane, following homogeneous Poisson point process with density $\lambda_b$. Similarly, user coordinates $\Phi_u$ follow homogeneous Poisson point process with density $\lambda_u$, independent of $\Phi_b$. Users are associated with their nearest BSs, which correspondingly forms BS coverage regions whose boundaries comprise a two-dimensional Voronoi tessellation \cite{StoyanBook:StochasticGeometry:1995}. Each BS is tuned off when its coverage, a Voronoi cell, is empty of serving users, otherwise transmitting with unity power. For a transmitted signal, we consider path loss attenuation with the exponent $\alpha>2$ and Rayleigh fading. 



\begin{figure} 
\includegraphics[width=8.7cm]{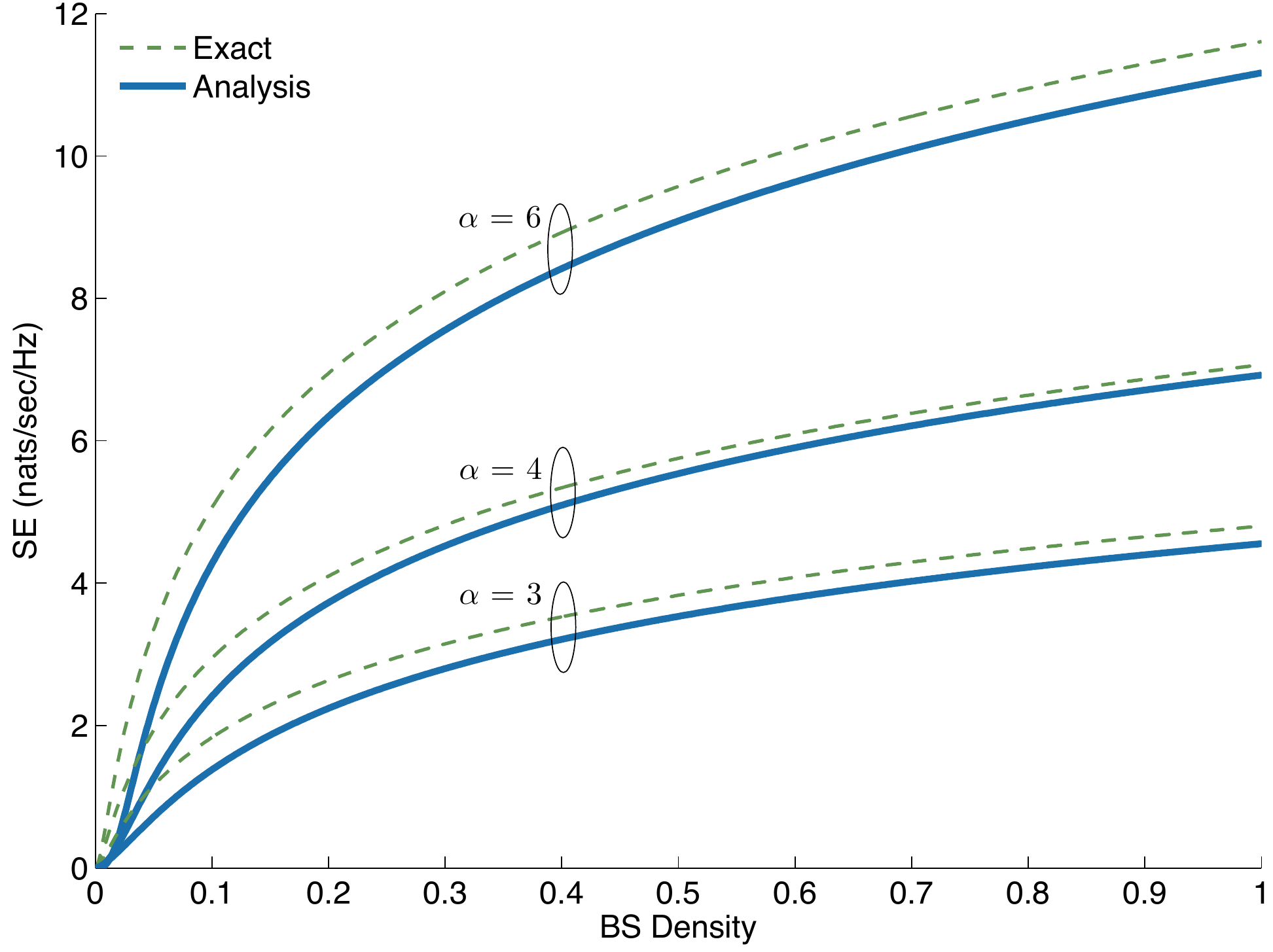}  
	\caption{Exact SE (equation \eqref{Eq:Exact}) versus its analytically approximated curve (equation \eqref{Eq:UDN} in Proposition 1) for path loss exponents $\alpha=3$, $4$, and $6$ where user density $\lambda_u=0.02$. Compared to the exact SE when $\alpha=4$, the analytic value achieves: 81.75\%, 90.9\%, and 97.96\% for BS density $\lambda_b=0.1$, $0.2$, and $1$ respectively.}
	\label{Fig:CapApprox}
\end{figure}

\subsection{Average Spectral Efficiency} \label{Sect:Cap_Approx}
This section aims at providing a closed-form SE, to be utilized for profit maximization in Section \ref{Sect:ProfitMax}. To this end, we consider interference-limited regime neglecting noise, and define SE $\gamma$ as ergodic rate per user for unity spectrum amount, $  \E\log\[1 + \SIR \]$ in units of nats/sec/Hz (1bit $\approx$ 0.693 nats) where $\SIR$ denotes signal-to-interference ratio. For brevity, we neglect multiple user access, to be considered in Section \ref{Sect:Cap_Approx_MA}.

According to Proposition 1 in \cite{Yu2011}, a BS's turned-on (non-empty) probability $p_a$ is given as $1-\( 1+3.5^{-1}\frac{\lambda_u}{\lambda_b} \)^{-3.5}$.
Applying this to the equation (16) in \cite{Andrews:2011bg} provides $\gamma$ as:
\begin{equation} 
\gamma =   \int_0^\infty \l[ 1 + \rho_t  \l(e^t - 1 \r)^{\frac{2}{\alpha}} p_a    \r]^{-1}\hspace{-7pt} dt \label{Eq:Exact}
\end{equation}
where $\rho_t :=  \int_{(e^t-1)^{-\frac{2}{\alpha}}}^\infty 1/\l(1 + u^{\frac{\alpha}{2}}\r)du$.\vspace{2pt}\\ 

For a sparse ($\lambda_b \ll \lambda_u$) cellular network when $p_a \approx 1$, the equation \eqref{Eq:Exact} is in accord with the result in \cite{Andrews:2011bg}. For explanatory convenience, let $\gamma_\alpha$ hereafter denote the SE without multiple access in a sparse network, given as
\begin{equation}
\gamma_\alpha = \int_0^\infty \[ 1 + \rho_t (e^t-1)^{\frac{\alpha}{2}} \]^{-1} dt. \label{Eq:Sparse}
\end{equation}

This shows $\gamma_\alpha$ is independent of $\lambda_b$. It implies, in other words, BS densification does not yield any gain in SE under sparse environment since its desired signal power improvement is on the same order of amount as the aggregate interference increase, cancelled out each other at calculating $\SIR$.

For an ultra-dense ($\lambda_b \gg \lambda_u$) network when $p_a \approx \frac{\lambda_u}{\lambda_b}$ by using Taylor expansion, the equation \eqref{Eq:Exact} hardly captures the relationship between $\gamma$ and $\lambda_b$ due to the complicated double integrations therein. We take a detour this problem by deriving the closed-form approximation of $\gamma$ in the following proposition.

\begin{proposition}\emph{
\emph{(Approximated SE in Ultra-Dense Networks)} SE in an ultra-dense downlink cellular network is given as:
\begin{eqnarray}
\gamma&\gtrsim&  \log\l[ 1 + \l( \frac{\lambda_b}{\rho_0 \lambda_u} \r)^{\frac{\alpha}{2}} \r] \label{Eq:UDN}
\end{eqnarray} 
where $\rho_0 :=  \int_{0}^\infty 1/\l(1 + u^{\frac{\alpha}{2}}\r)du$.\\
\begin{proof} See Appendix.
\end{proof}}
\end{proposition}
Interestingly, this simple expression shows that $\gamma$ is a logarithmic function of $\lambda_b$. In other words, \emph{BS densification does increase the SE whilst yielding diminishing returns} under ultra-dense environment. 

A stochastic geometric point of view interprets this phenomenon as follows. At a typical user, increasing $\lambda_b$ shrinks each BS's coverage, simultaneously yielding: (i) the shortened distance to his associated BS and (ii) increased the number of empty cells (or turned-off BSs). Focusing firstly on the former, the shortened distance in the order ${\lambda_b}^{-\frac{1}{2}}$ (see the average distance is $1/(2\sqrt{\lambda_b})$ in \cite{Haenggi:05}) yields the received signal power increase from the associated BS, in the order ${\lambda_b}^{\frac{\alpha}{2}}$. For the latter in an ultra-dense scenario, it makes almost all BSs turned-off except for the ones serving users in their infinitesimal coverage regions. Consequently, the interfering BS locations (or interferer density) converge to the users' (or $\lambda_u$), delimiting the quantity of interference. Combining these results leads to the ever-increasing $\SIR$ in the order ${\lambda_b}^\frac{\alpha}{2}$, resulting in the logarithmic SE increase. 

Fig. \ref{Fig:CapApprox} visually validates the tightness of the value in Proposition 1 for different $\alpha$'s. When $\alpha=4$, for instance, the difference between \eqref{Eq:Exact} and \eqref{Eq:UDN} is less than $15\%$ for $\lambda_b \geq 6 \lambda_u$. Thus, we hereafter regard \eqref{Eq:UDN} as the approximation of $\eqref{Eq:Exact}$. 

Additionally, it is worth mentioning that this result shows the BS ultra-densification gain in SE. Comparing to $\gamma_\alpha$ in a sparse scenario, ultra-densification provides the SE gain: 162\%, 250\%, and 464\% for $\lambda_b=0.1$, $0.2$, and $1$ respectively when $\lambda_u=0.02$.

\subsection{Average Spectral Efficiency with Multiple Access} \label{Sect:Cap_Approx_MA}
Now we turn our attention to multiple access of users with a fixed amount of spectrum. So far we have considered a BS serves all users in its coverage. Instead, we henceforth consider each BS serves at most a single user at a given time, who is selected according to a uniformly random scheduler \cite{TseBook:FundamaentalsWC:2005}. According to Proposition 2 in \cite{Yu2011}, a typical user's selection probability by the scheduler for a sparse network is $\frac{\lambda_b}{\lambda_u}$, and for an ultra-dense network is $1$. Applying these results to the equations \eqref{Eq:Sparse} and \eqref{Eq:UDN} yields the following corollary.

\begin{corollary}\emph{\emph{(SE with Multiple Access)} SE with a uniformly random scheduler in a sparse or ultra-dense downlink cellular network is given as follows.
\begin{eqnarray}
\emph{Sparse:}\;\;\; \gamma &\hspace{-3pt} \approx& \hspace{-3pt} \frac{\lambda_b}{\lambda_u} \gamma_\alpha \hspace{10pt}\label{Eq:SparsewMA}\\
\emph{Ultra-Dense:}\;\;\; \gamma &\hspace{-3pt} \approx& \hspace{-3pt}  \log\l[ 1 + \l( \frac{\lambda_b}{\rho_0 \lambda_u} \r)^{\frac{\alpha}{2}} \r] \label{Eq:UDNwMA}
\end{eqnarray}
}\end{corollary}

In a sparse scenario, increasing $\lambda_b$ alleviates multiple access congestion, and thereby linearly increases SE in spite of a constant $\gamma_\alpha$ independent of $\lambda_b$ in \eqref{Eq:SparsewMA}.

In an ultra-dense scenario, on the other hand, reducing the access congestion along with BS densification does not ameliorate SE. That is because multiple access of users barely occurs in the ultra dense network where almost all BSs have at most a single user within their coverages. This results in \eqref{Eq:UDNwMA} being equivalent to \eqref{Eq:UDN}. The following section utilizes these results so as to explore the economic impact of BS densification.

\begin{figure}
\centering
	\includegraphics[width=8.7cm]{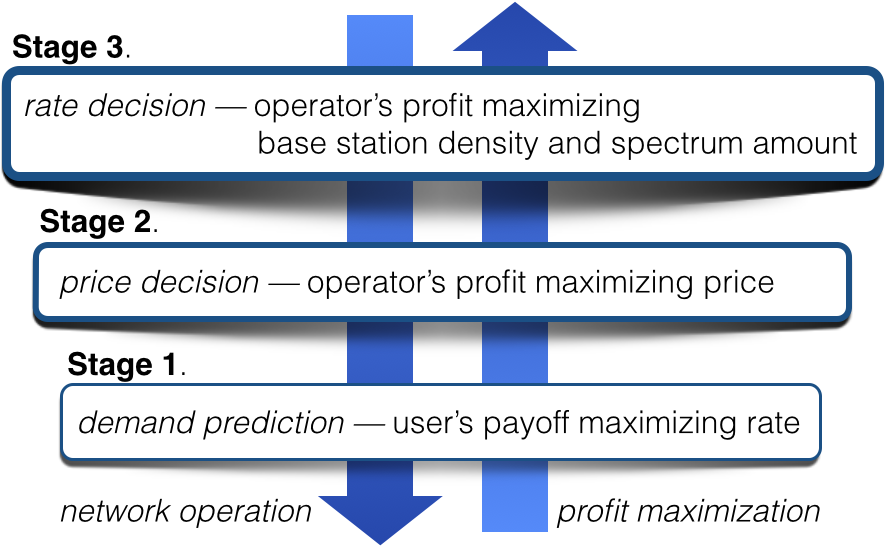}  
	\caption{Three-stage profit maximization. In order to maximize profit, an operator predicts user demand at Stage 1, determines price at Stage 2, and decides operating BS density and spectrum amount at Stage 3. After the final decision, the network operates in the reverse direction, from Stage 3 to 1.} \label{Fig:3Stages}
\end{figure}

\section{User Demand-Aware Profit Maximization} \label{Sect:ProfitMax}
In this section, our objective is to maximize a network operator's profit when user demand varies, caused by the changes in the number of users $\lambda_u$ and/or each user's rate sensitivity $b$. The operator is able to cope with these demand changes by adjusting his operating BS density $\lambda_b$ and spectrum amount $W$ as well as price per nats/sec (or per nat for unity time) $p$. Since the profit is a joint function of user demand, price, average rate, and its operating cost, the operator's profit maximizing decision problem to clarify the adjustments in $\lambda_b$, $W$, and $p$ is not trivial, being of our interest. To be more specific, we firstly predict average per-user demand $\bar{X}$, and then determine optimal $\lambda_b$ and $W$ so as to maximize the average profit per unit area, formulated as:
\begin{eqnarray*}
\texttt{(P1): } \underset{\lambda_b, W, p}{\texttt{Maximize}} && p \lambda_u \bar{X} - \(c_b \lambda_b + c_w W  \) \\
\texttt{subject to} && \bar{X} \leq W \gamma
\end{eqnarray*}
where $c_b$ and $c_w$ respectively denote BS and unit spectrum operating costs per unit area. In an operator's perspective, we divide this profit maximization problem into three sequential stages: Stage 1. user demand prediction; Stage 2. price decision; and Stage 3. the decision on BS density and spectrum amount, resulting in average rate to be supplied. Fig. \ref{Fig:3Stages} elucidates these subdivided problems and their solving direction as well as the direction of network operation for given solutions of the problem.

\subsection{User Demand Prediction (Stages 1 and 2)} \label{Sect:DemandModel}
For the purpose of predicting user demand at Stage 1, consider a typical user's payoff $U$ having the following characteristics: logarithmically increasing with downloading rate \cite{Shaikh:QoEUsrNet:2010}, and linearly decreasing with cost under usage-based pricing \cite{SenJoeETAL:IncTstDatSurv:2012}. Let $X$ denote average rate per user, and define $\theta$ as a user's willingness-to-pay, assumed to be uniformly distributed from $0$ to $b$, $\forall b>p$. We hereafter interpret the maximum willingness-to-pay, $b$, as the user's rate sensitivity. Correspondingly, we represent $U$ as:
\begin{equation} 
U = [\theta \log\l( 1 + X \r) - p X]^+.
\end{equation}

Consider users try to maximize their payoffs. Since $U$ is a concave function of $X$, applying the first order necessary condition and taking average over $\theta$ yield the payoff maximizing average rate per user (or average demand per user) $\bar{X}$ as follows.
\begin{equation}
\bar{X} = \frac{(b - p)^2}{2 b p} \label{Eq:AvgDemand}
\end{equation}

\begin{figure*}
\centering
 	\subfigure[For user density $\lambda_u$ for $b=10$]{\includegraphics[width=8.7cm]{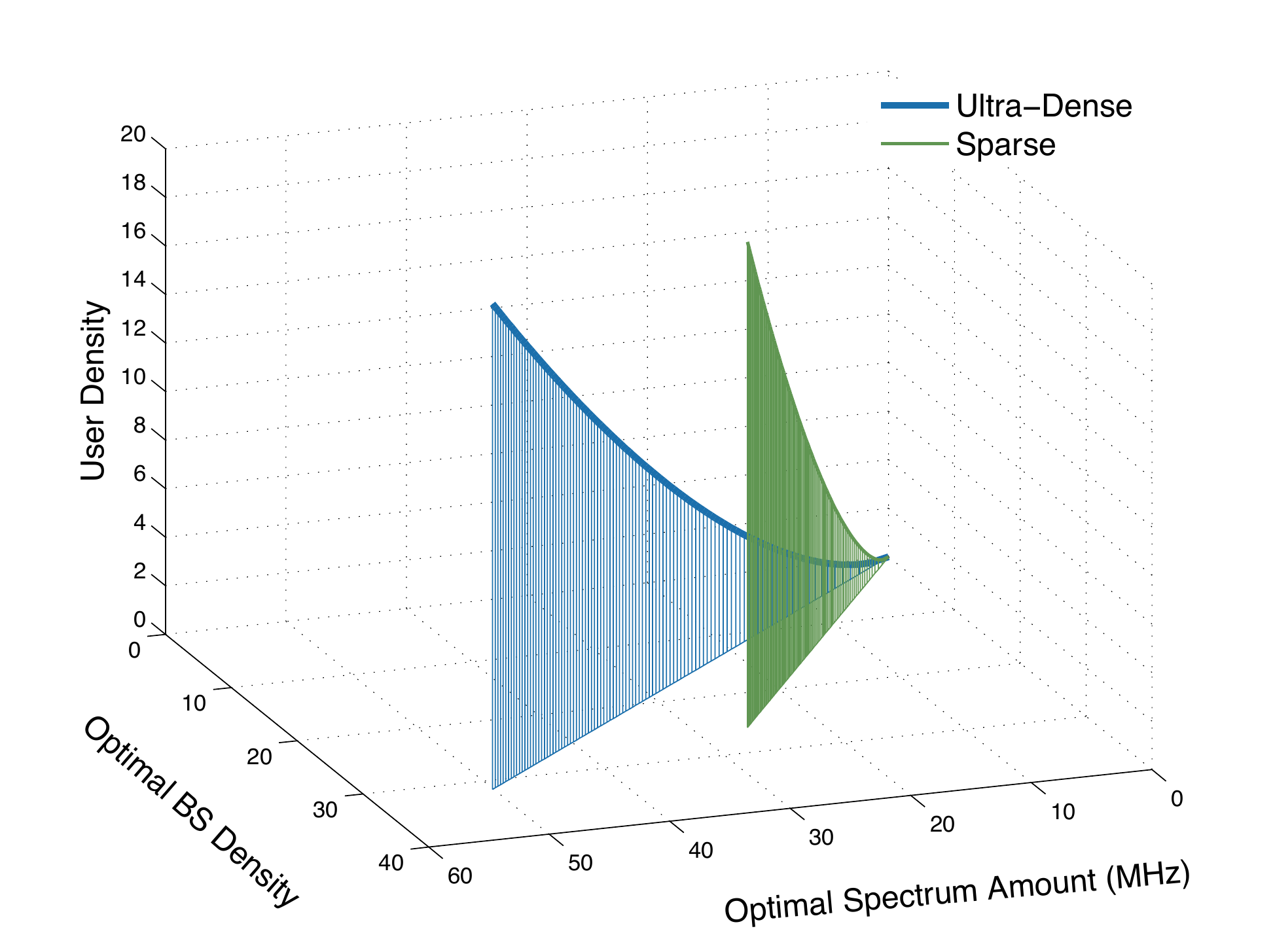}  }\label{Fig:Approx_LowDen} 
 	\subfigure[For per-user rate sensitivity $b$ for $\lambda_u=5$]{\includegraphics[width=8.7cm]{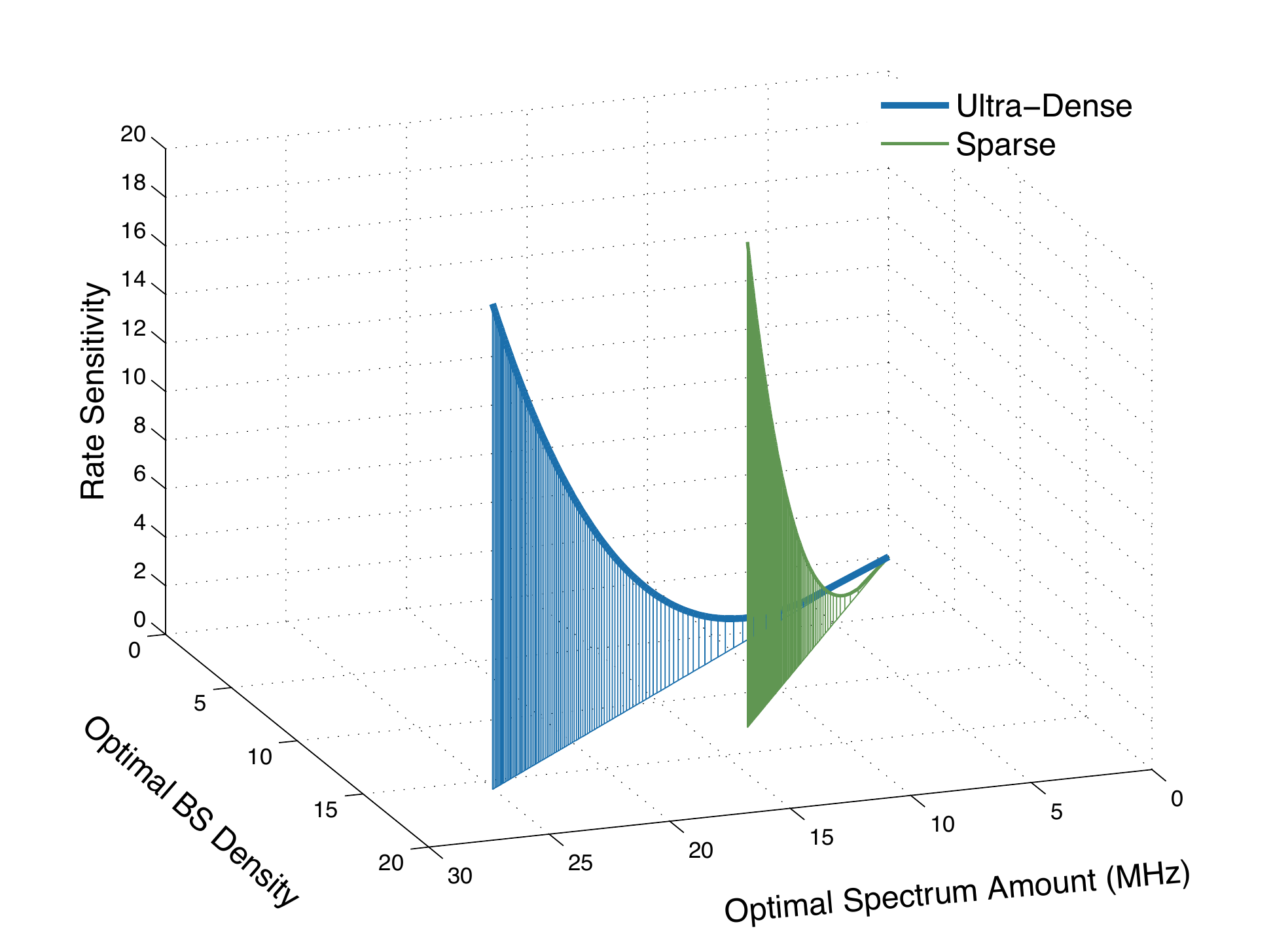}  }\label{Fig:Approx_HighDen}
	\caption{Profit optimal BS density $\lambda_b^*$ and spectrum amount $W^*$ ($\alpha=4$, $c_b=c_w=0.1$).}
	\label{Fig:OptWdensB}
\end{figure*}

Applying the result to the profit function in \texttt{(P1)} reveals the profit decreases with $p$. This intuitively indicates attracting more users by reducing $p$ yields higher profit than increasing $p$. Since $\bar{X}$ is also a decreasing function of $p$, the profit maximizing price $p^*$ at Stage 2 is the price when the equality holds at the constraint in \texttt{(P1)}, resulting in
\begin{eqnarray}
p^* &=& b(1 + W\gamma)\[ 1- \( 1 - \frac{1}{(1 + W \gamma)^2}\)^{\frac{1}{2}}\] \nonumber \\ 
&\overset{(a)}{\approx}& \frac{b}{2(1 + W \gamma)} \label{Eq:OptPrice}
\end{eqnarray}
where (a) follows from Taylor expansion for large $W\gamma$.

\subsection{Profit Optimal BS Density and Spectrum Amount (Stage 3)} \label{Sect:Opt}
This subsection aims at deriving optimal BS density $\lambda_b^*$ and spectrum amount $W^*$ in closed forms so that their resultant average rate is provided to a typical user while maximizing profit.

Exploiting $p^*$, the equation \eqref{Eq:OptPrice} in Section \ref{Sect:DemandModel}, modifies \texttt{(P1)} for Stage 3 as follows.
\begin{eqnarray*}
\texttt{(P2): } \underset{\lambda_b, W}{\texttt{Maximize}} && \frac{\lambda_u b}{2} \(1+\frac{1}{W \gamma}\)^{-1} - \(  c_b \lambda_b + c_w W \) 
\end{eqnarray*}
For sparse cellular networks, applying \eqref{Eq:SparsewMA} to $\gamma$ of \texttt{(P2)} yields the following profit maximization problem.
\begin{eqnarray*}
\texttt{(P3): } \underset{  \lambda_b, W}{\texttt{Maximize}} && \hspace{-15pt} \frac{\lambda_u b}{2}\(1 + \frac{\lambda_u}{W \lambda_b \gamma_\alpha}\)^{-1} - \(c_b \lambda_b + c_w W  \) 
\end{eqnarray*}
For ultra-dense networks, in the same manner, applying \eqref{Eq:UDNwMA} to $\gamma$ in \texttt{(P2)} leads to the following profit maximization problem.
\begin{eqnarray*}
\texttt{(P4): } &&\\  
&& \hspace{-40pt} \underset{\lambda_b, W}{\texttt{Maximize}}  \; \frac{\lambda_u b}{2} \l\{1 + \( W \log\[ 1 + \(\frac{\lambda_b}{\rho_0\lambda_u}\)^{\frac{\alpha}{2}} \] \r)^{-1} \r\}^{-1}\\
&& \hspace{10pt} - \( c_b \lambda_b + c_w W \) 
\end{eqnarray*}
Solving \texttt{(P3)} and \texttt{(P4)} provides profit maximizing $\lambda_b^*$ and $W^*$ of sparse and ultra-dense networks respectively, as provided in the following proposition.

\begin{proposition} (Optimal BS Density and Spectrum Amount) \emph{ Profit optimal operating BS density and spectrum amount in a sparse or ultra-dense downlink cellular network are given as:
\begin{eqnarray}
\emph{Sparse:} \hspace{-15pt}&\hspace{-3pt}&\left\{\begin{array}{lll}
             \lambda_b^*  &\hspace{-3pt} =& \hspace{-3pt}  \[ \frac{b c_w }{2 \gamma_\alpha } \(\frac{\lambda_u}{c_b}\)^2  \]^{\frac{1}{3}} \\ \vspace{-5pt}\\
              W^*  \hspace{-3pt}&\hspace{-3pt} =& \hspace{-3pt}  \[ \frac{b c_b }{2 \gamma_\alpha } \(\frac{\lambda_u}{c_w}\)^2  \]^{\frac{1}{3}}
            \end{array}\right. \label{Eq:OptSparse}\\
\emph{Ultra-Dense:}\hspace{-15pt}&\hspace{-3pt}& \left\{\begin{array}{lll}
             \lambda_b^*  &\hspace{-3pt} \approx& \hspace{-3pt}  \[ \( \frac{\alpha}{2^{2.5} c_b} \)^8 (b c_w)^4 {\rho_0}^\alpha {\lambda_u}^{\alpha + 4} \]^{\frac{1}{\alpha +8}} \\ \vspace{-5pt}\\
              W^*  \hspace{-3pt}&\hspace{-3pt} \approx& \hspace{-3pt}  \[ 2^{2( \alpha -2)} \frac{{c_b}^\alpha}{{c_w}^{\alpha + 4}} b^4 {\rho_0}^\alpha {\lambda_u}^{\alpha + 4} \]^{\frac{1}{\alpha + 8}}.
            \end{array}\right.            \label{Eq:OptUDN}
\end{eqnarray} 
\begin{proof} See Appendix.
\end{proof}
}\end{proposition}

We interpret the above results in the following perspectives: 1) unit operating costs $c_b$ and $c_w$ and the resultant profit optimal operating costs $c_b \lambda^*$ and $c_w W^*$ and 2) user demand comprising per-user rate sensitivity $b$ and user density $\lambda_u$.

\begin{figure*}
\centering
 	\subfigure[For user density $\lambda_u$ for $b=10$]{\includegraphics[width=8.7cm]{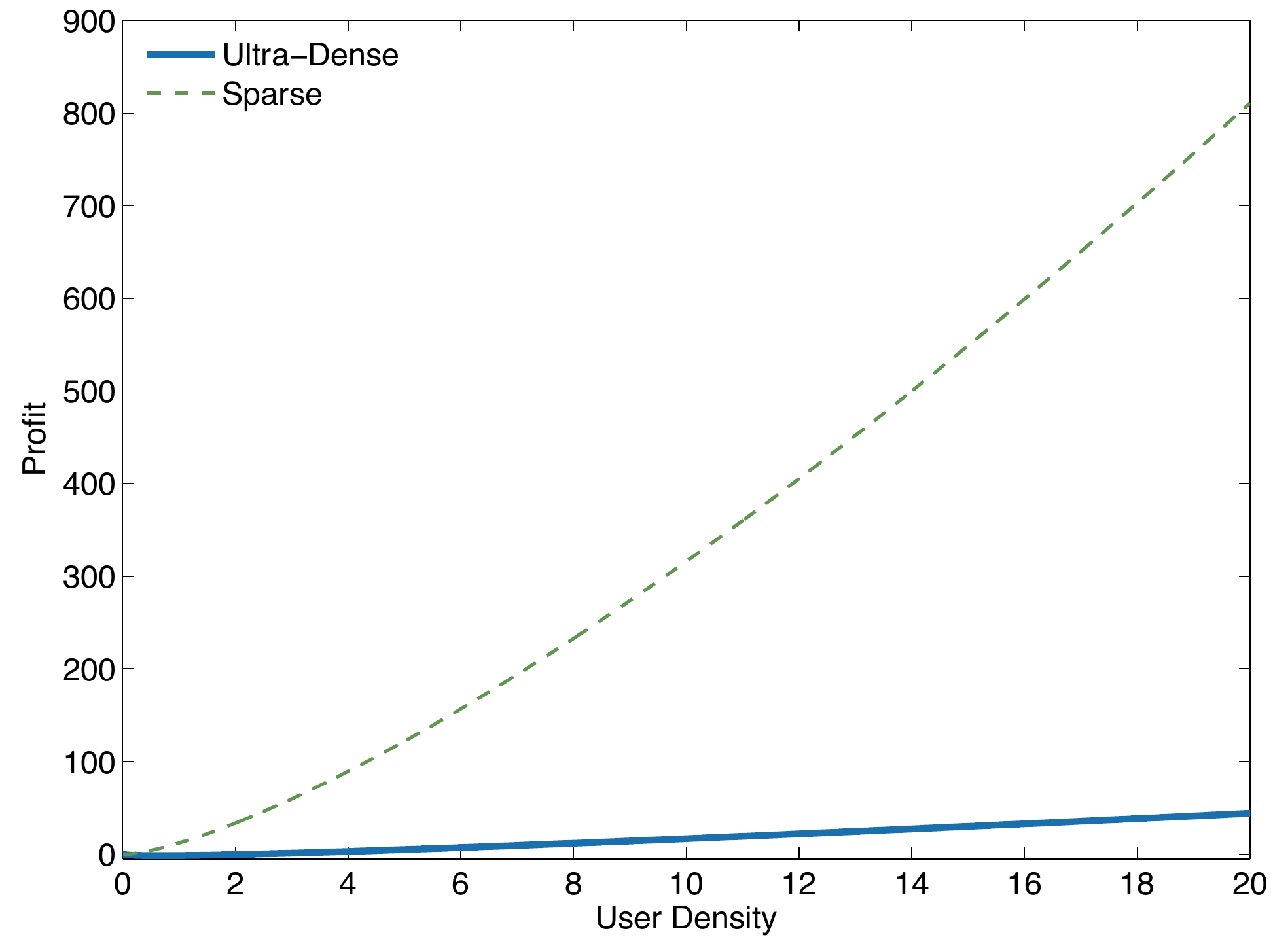}  }
 	\subfigure[For per-user rate sensitivity $b$ for $\lambda_u=1$]{\includegraphics[width=8.7cm]{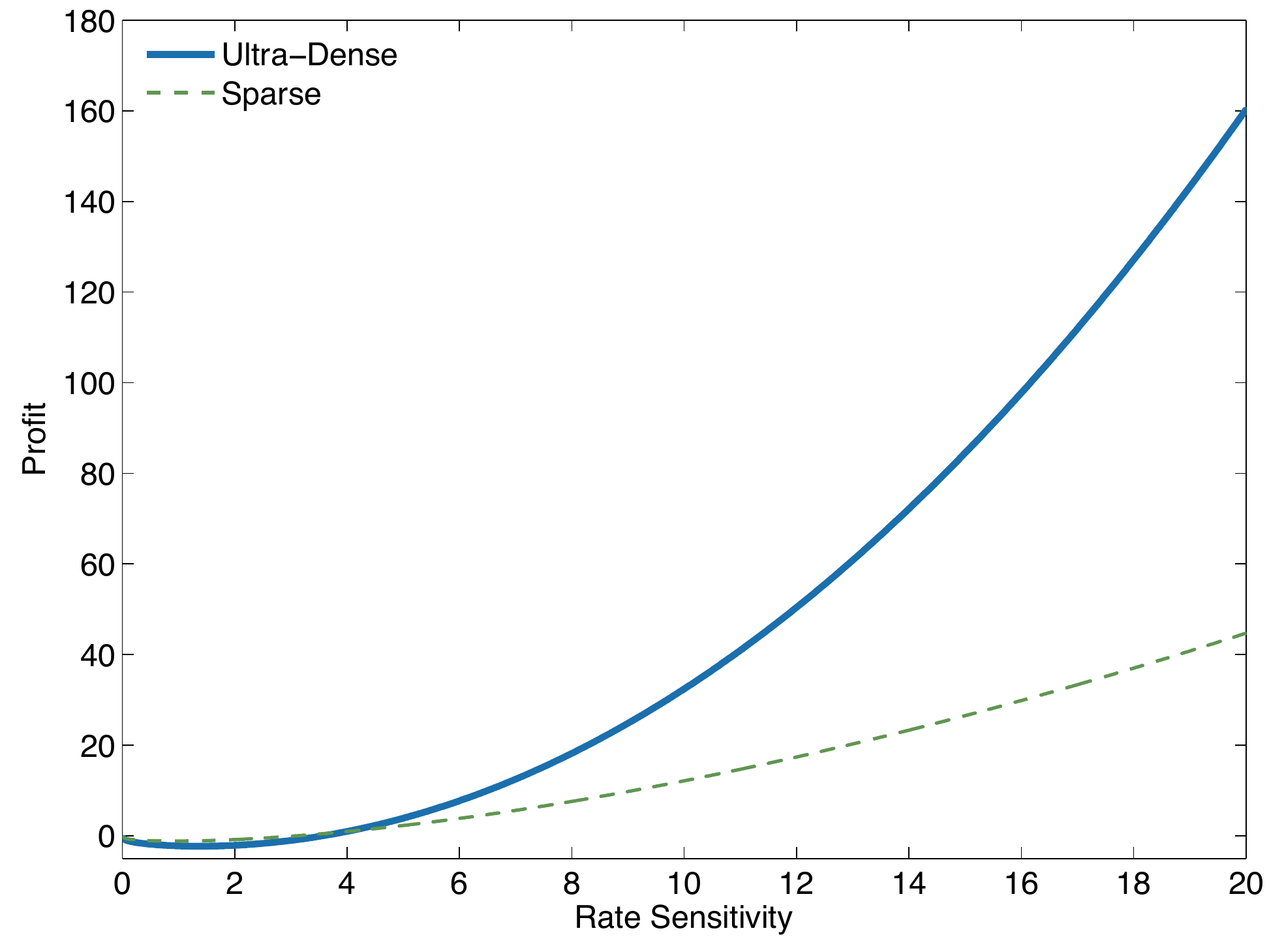}  }
	\caption{Maximized profit with optimal BS density $\lambda_b^*$ and spectrum amount $W^*$ ($\alpha=4$, $c_b=c_w=0.1$).}
	\label{Fig:OptWdensB}
\end{figure*}

\subsubsection{Effect of Operating Cost}
Increasing unit BS operating cost $c_b$ leads to investing more in spectrum as the BS substitute, captured by increasing $W^*$, and vice versa for unit spectrum operating cost $c_w$ increase.

Multiplying these unit operating costs by \eqref{Eq:OptSparse} and \eqref{Eq:OptUDN} yields the following profit optimal operating BS/spectrum cost ratio.

\begin{corollary}\emph{\emph{(Optimal Cost Ratio)} Profit maximizing ratio of BS and spectrum operating costs in a sparse or ultra-dense downlink cellular network is given as:
\begin{eqnarray}
\emph{Sparse: }\quad \frac{c_b \lambda_b^*}{c_w W^*}&=& 1 \\ \nonumber\\
\emph{Ultra-Dense:}\quad \frac{c_b \lambda_b^*}{c_w W^*} &\approx& 2^{-2} \alpha^{\frac{8}{\alpha +8}}.
\end{eqnarray}
}\end{corollary}

In a sparse network, the operator should invest in BS operating cost as much as the spectrum cost since BS density and spectrum amount equally affect average rate (see \eqref{Eq:SparsewMA} with the spectrum amount $W$, shown in \texttt{(P3)}). In an ultra-dense network, on the other hand, BS density less affects average rate than spectrum amount due to the densification's logarithmic impact on average rate. This leads to the investment strategy that BS operating cost should be less than the spectrum cost, depending on $\alpha$ (note $ 0.43 \leq 2^{-2}\alpha^{\frac{8}{\alpha+8}} \lesssim 0.71$). The straight lines on the BS density-and-spectrum planes (bottom) in Fig. 3 illustrate such operating cost ratios.

\subsubsection{Effect of User Demand}
Both profit optimal BS density and spectrum amount increase as user demand grows, but the optimal value increments incurred by user density are higher than the values by per-user rate sensitivity (see the exponents of $\lambda_u$ and b in \eqref{Eq:OptSparse} and \eqref{Eq:OptUDN}, visualized in Fig. 3). 

While rate sensitivity growth solely increases user demand, user density growth not only increases the demand but also decreases average rate due to: (i) incurring more multiple access congestion in a sparse network or (ii) generating more interference in an ultra-dense network (see the discussion after Proposition 1 in Section \ref{Sect:Cap_Approx}). Consequently, increasing user density requires more BS density and/or spectrum amount in order to keep up with the demand growth as well as to recover the average rate decline.

Such different impacts of user density and per-user rate sensitivity furthermore affect the network profitability. For user density growth, Fig. 4(a) shows the profit of an ultra-dense network increases less than that of a sparse network. An ultra-dense network requires much more BS density and spectrum amount to compensate the interference generation caused by user density growth, and it engenders too much cost increase worsening the network profitability. For this type user demand increase, BS ultra-densification is not preferable.

For per-user rate sensitivity growth, in contrast, Fig. 4(b) depicts the profit of an ultra-dense network increases more than that of a sparse network thanks to the network's delimited interference, promoting ultra-dense BS deployment.

\section{Discussion}
In this paper we have derived a closed-form relationship between BS density and SE in an ultra-dense cellular network. The SE is shown to be a logarithmic function of BS density as the density grows. This closed-form SE expression was used to derive closed-form solutions for the optimal operating BS density and spectrum amount, for the traditional spectrum licensing (auction) case. This expression could aid the operator in his decision if he should invest more in BS densification or bid for more spectrum for his network.  Our results reveal some fundamentally unique characteristics of the ultra-dense network deployment, e.g. that the number of users has a larger impact on the optimal network configuration than each user's sensitivity to his downloading rate. Further, in our simplified model we see that the network operator should maintain a reasonable balance in investment, allocating about equal amounts to spectrum and BS deployment.

To the best of our knowledge, this paper is the first to simultaneously specify not only random spatial locations of both users and BSs but also each user's demand model, bridging the gap between stochastic geometric and network economic analysis.

A weakness of the study is the very simple, homogeneous propagation model, i.e. using a constant path loss exponent. The BSs are typically in the same room as the users in line-of-sight (LOS) conditions, whereas the interfering BSs are behind walls, i.e.  in non-LOS conditions, creating a better SIR. Further work should therefore involve a two-slope model, describing the LOS and non-LOS cases and giving a more direct relationship to the physical environment. Further extension to this work could also include milimeter-wave systems where non-LOS signals are very weak. In addition, it does not seem likely that future short range systems will use traditionally licensed spectrum. Extending the economic analysis to spectrum sharing paradigms is another interesting avenue for future research.

\section*{Acknowledgement} \small This research was supported by the International Research \& Development Program of the National Research Foundation of Korea (NRF) funded by the Ministry of Education, Science and Technology (MEST) of Korea (Grant number: 2012K1A3A1A26034281). \normalsize

\appendix

\subsection{Proof of Proposition 1}
Let $a$ denote $\rho_0 \lambda_u/\lambda_b$. Since $\rho_0 \geq \rho_t$, $\gamma$ is lower bounded as:
\begin{eqnarray}
\gamma &\geq& \int_{t>0} \[ 1 + a (e^t - 1)^{\frac{2}{\alpha}} \]^{-1} dt \nonumber \\
&\overset{(a)}{\geq}& \int_{t>0} \[ 1-a (e^t - 1)^{\frac{2}{\alpha}} \]^+ dt\\
&=& \log\( 1 +a^{-\frac{\alpha}{2}}\) + a \pi \csc\( \frac{2 \pi}{\alpha} \) \nonumber\\
&&\hspace{-5pt} -\frac{\alpha }{2\( 1 + a^{\frac{\alpha}{2}}\)} \underbrace{ \, _2F_1\left(1,1;1-\frac{2}{\alpha};1-\frac{1}{a^{\frac{\alpha}{2}}+1}\right)}_{(b)} 
 \label{Eq:Prop1Pf}
\end{eqnarray}
where $(a)$ follows from Taylor expansion, Gaussian hypergeometric function $\, _2F_1(a, b; c; z) : = \sum _{k=0}^{\infty } \frac{z^k a^{(k)} b^{(k)} }{k! c^{(k)}}$, and $x^{(k)}$ rising factorial. The function $(b)$ monotonically increases with $a$ for all $\alpha$, having unity minimum value at $a=0$. For $a \ll 1$ (or $\lambda_b \gg \lambda_u$), therefore all terms in \eqref{Eq:Prop1Pf} except $\log(1 + a^{-\frac{\alpha}{2}})$ become negligible, completing the proof. \hfill $\blacksquare$

\subsection{Proof of Proposition 2}
Consider a sparse network. For sufficiently large average rate (or small $\lambda_u/\(W \lambda_b \gamma_\alpha\)$), applying Taylor expansion to the objective function in \texttt{(P3)} leads to the following problem.
\begin{eqnarray*}
\texttt{(P3.1): } \underset{\lambda_b, W,}{\texttt{Maximize}} && \hspace{-15pt} \frac{\lambda_u b}{2}\(1 - \frac{\lambda_u}{W \lambda_b \gamma_\alpha}\) - \(c_b \lambda_b + c_w W  \) 
\end{eqnarray*}
The profit function is concave with respect to both $\lambda_b$ and $W$, so it has a unique maximum result. Exploiting the first order necessary condition yields
\begin{eqnarray}
\lambda_b^* &=& \[ \frac{b}{2 \gamma_\alpha c_b W^*} \]^{\frac{1}{2}} \lambda_u \quad \text{and} \label{Eq:PfProp3densB}\\
W^*&=&\[ \frac{b}{2 \gamma_\alpha c_w \lambda_b^*} \]^{\frac{1}{2}}\lambda_u.  \label{Eq:PfProp3W}
\end{eqnarray}
Applying \eqref{Eq:PfProp3W} to \eqref{Eq:PfProp3densB} proves the result.

Next, consider an ultra-dense network. Since the logarithmic function in \texttt{(P4)} is not appropriate for deriving solutions in closed forms, we resort to considering its lower bound $\frac{1}{\log\( 1 + x\)} \geq x^{-\frac{1}{2}}$ as the approximation that is tight for large $x$ (or $\lambda_b \gg \lambda_u$). Applying Taylor expansion provides the formulation as shown below.
\begin{eqnarray*}
\texttt{(P4.1): }  &&\\
&& \hspace{-50pt} \underset{ \lambda_b, W}{\texttt{Maximize}} \frac{\lambda_u b}{2}\l\{1 -\frac{1}{W} \(\frac{\rho_0\lambda_u}{\lambda_b}\)^{\frac{\alpha}{2}} \r\} - \(  c_b \lambda_b + c_w W\) 
\end{eqnarray*}
In the same way as the sparse network, exploiting the first order necessary condition leads to 
\begin{eqnarray}
\lambda_b^* &=& \[ \frac{\alpha b {\rho_0}^{\frac{\alpha}{4}}}{2^3 c_b W^*} \]^{\frac{1}{\frac{\alpha}{4}+1}} \lambda_u \quad \text{and} \label{Eq:PfProp3densBudn}\\
W^*&=&\[ \frac{b {\lambda_u}^{\frac{\alpha}{4}+1}  }{2 c_w}\( \frac{\rho_0}{{\lambda_b}^*}\)^{\frac{\alpha}{4}} \]^{\frac{1}{2}}.  \label{Eq:PfProp3Wudn}
\end{eqnarray}
Applying \eqref{Eq:PfProp3Wudn} to \eqref{Eq:PfProp3densBudn} finalizes the proof. \hfill $\blacksquare$

\end{document}